\documentclass[10pt]{article}

\usepackage[]{graphicx}
\RequirePackage{hyperref}

\graphicspath{{img/}{fig3/}}

\usepackage[left=2cm,top=2cm, right=2cm, nohead,nofoot]{geometry}

\usepackage{placeins}
\usepackage[latin1]{inputenc}
\usepackage{url}
\usepackage{hyperref}
\usepackage{color}
\usepackage{array,booktabs}
\usepackage{color}

\title{Predicting human mobility through the assimilation of social media traces into mobility models}

\author{Mariano G. Beir\'{o}$^1$ \qquad Andr\'{e} Panisson$^1$ \qquad Michele Tizzoni$^1$ \qquad Ciro Cattuto$^1$\vspace{0.4cm}\\$^1$ ISI Foundation, Turin, Italy}

\date{}
\begin{document}
\maketitle

\begin{abstract}

Predicting human mobility flows at different spatial scales is challenged by the heterogeneity of individual trajectories and the multi-scale nature of transportation networks. 
As vast amounts of digital traces of human behaviour become available, an opportunity arises to improve mobility models by integrating into them proxy data on mobility collected by a variety of digital platforms and location-aware services.
Here we propose a hybrid model of human mobility that integrates a large-scale publicly available dataset from a popular photo-sharing system with the classical gravity model, under a stacked regression procedure. 
We validate the performance and generalizability of our approach using two ground-truth datasets on air travel and daily commuting in the United States: using two different cross-validation schemes we show that the hybrid model affords enhanced mobility prediction at both spatial scales.

\end{abstract}

\section{Introduction}

Modelling and understanding human mobility patterns at different spatial scales and aggregation levels - from single individuals to population displacements  - is an important research topic because of a vast number of applications, ranging from urban and transportation planning~\cite{roth2011,lenormand2014} and resource allocation~\cite{song2006evaluating, abou2013optimal} to the prediction of migration flows~\cite{ravenstein1885laws, simini2012universal} and epidemic spreading at local, regional or worldwide level~\cite{Wesolowski12102012, balcan2009multiscale, merler2010, marguta_2015}.

In the last few years, a significant research effort has been made to understand human mobility patterns, both in the laws governing individual human trajectories~\cite{gonzalez2008, rhee2011} and collective movements~\cite{noulas, simini2012universal, hawelka2014}. 
In the latter case, the most extensively used models are the gravity model~\cite{zipf49, alonso1976theory} and the more recent radiation model~\cite{simini2012universal}. 
The gravity model assumes that the amount of people travelling between two locations is directly proportional to some power of their populations, and decays as some power of the distance between them. 
Instead, the radiation model considers human movements as diffusion processes that depend on the population distribution over the space, reproducing Stouffer's theory of intervening opportunities~\cite{stouffer}. 
Both models are static and require some information in order to be adjusted: in the gravity model, parameters are fitted using real mobility data, provided by an independent source; the radiation model, in its original formulation, is parameter-free, but it requires accurate knowledge of the spatial population distribution.
Both modelling approaches have been extensively tested, showing advantages and limitations. 
The gravity model has been successfully used to describe highway flows~\cite{highways_2008}, air-travel~\cite{grosche2007gravity,liu2014uncovering}, commuting~\cite{balcan2009multiscale} and mobile phone calls between cities~\cite{krings_calls}. However, it has some relevant limitations, as the availability of data for calibration and the lack of a first principle derivation~\cite{simini2012universal, masucci2013}. 
On the other hand, the radiation model offers very good predictions for commuting patterns between U.S. counties using only population data, but its applicability at different spatial scales has been debated since it does not succeed in capturing commuting inside urban or metropolitan areas~\cite{liang2013unraveling, masucci2013, yang2014limits}, and it has never been used to model long distance travel patterns either.

The limitations of these models suggest that the quality of their results can be largely improved if they are supported by additional data~\cite{truscott2012,yang2014limits}.
In fact, several works have analyzed records from mobile phone companies to study individual~\cite{gonzalez2008} and collective mobility~\cite{cal_2011, palchykov2014, alexander2015origin}, showing that it is possible to infer these flows from human activity. The mobility flows obtained in this way can be successfully used for the prediction of epidemic spreading~\cite{tizzoni2014, bengtsson2015}, as a proxy for the real, often inaccessible, mobility data. 

In this context, the large volumes of digital traces left by humans over the Internet allows for a better understanding of mobility processes, with immediate benefits. 
On the one hand, the increase in transport infrastructure during the last years and the dynamics of change of mobility patterns have brought the requirement of real-time modelling. People travel more, and travel patterns may change very fast, with important consequences for epidemic spreading and planning. A timely modelling of mobility processes might then allow for rapid interventions and for the design of emergency policies. 
On the other hand, though mobility data is usually available in many developed countries for airline transportation, train trips or commuting, in many underdeveloped countries this information is scarce or does not exist at all. The fact that mobility datasets are aggregated at a particular resolution level also constitutes a limitation for many potential studies. 

The scientific community recently recognized that one of the challenges in the modelling of social and epidemic processes is the assimilation of geolocalized data, and the construction of hybrid models combining metapopulation and network models with individual traces~\cite{riley2015five, gonccalves2015social}. 
Our approach to human mobility is in line with this perspective, as we analyze the effects of incorporating geolocalized traces from social media into the classical gravity model.

Social media platforms like Flickr~\cite{flickr_url}, Twitter~\cite{twitter_url} or Foursquare~\cite{foursquare_url} offer the possibility of georeferencing the content shared by users. Thus, they constitute a timely source of disaggregated, high-resolution spatio-temporal data on human mobility. 
The advantage of social media traces with respect to other sources of digital information is that they can be publicly accessed and at a very low cost. This approach have been taken by recent works in the literature, showing that mobility patterns can be successfully extracted from social media traces. Lenormand {\em et al.} used traces from Twitter to study highway and roadway transportation networks in Europe~\cite{lenormand2014}; Noulas {\em et al.} used a Foursquare dataset to analyze the link between user activity and place transitions~\cite{noulas2011}; Hawelka {\em et al.} modelled international travel of Twitter users by residence country~\cite{hawelka2014}; Lenormand {\em et al.} have also used Twitter traces to model commuting from home to work~\cite{lenormand2}; Grabowicz {\em et al.} studied the relation between human mobility and interactions using traces from different social networks~\cite{grabowicz2014}; Llorente et al.~\cite{llorente2015social} analyzed the mobility patterns in Spain using Twitter traces; Barchiesi et al.~\cite{Barchiesi150046} used Flickr data from $16,000$ individuals in the UK to model the flows between its $20$ largest cities, comparing their results with travel data obtained from surveys.

In this work we used a set of $18,000,000$ timestamped, georeferenced pictures from Flickr, taken by $40,000$ users in the U.S., which are part of the Yahoo Flickr Creative Commons 100M public dataset~\cite{thomee2015yfcc100m}. We processed the sequences of pictures belonging to each individual user in order to extract user trip paths at different resolution levels. Then, we used these emerging collective flow patterns to feed a learning model based on the gravity law.

Our main contribution is to design a data-driven hybrid model of human mobility, in which social media traces are combined with the classical gravity model under a machine learning approach, by training and cross-validating with real datasets. 
We evaluate the model for two different human activities and resolution scales: an air-travel network and a daily commuting network. Firstly, we show how individual traces can be adapted to these different resolution scales, by tessellating the space into adequate basins and filtering the correct individual flows. Secondly, we combine these traces with the gravity law and we fit the resulting hybrid model using a subset of the real data. Then, we evaluate the fitted model using the remaining part of the dataset. With a cross-validation procedure, we show that the hybrid model can be fitted using a small portion of the data as training set, to correctly predict the remaining mobility flows. In fact, we observe that the incorporation of Flickr traces into the gravity model improves its performance significantly, measured in terms of the determination coefficient. Our findings show that the Flickr traces are representative of the real human mobility and that they can be assimilated into a more theoretical model as the gravity model. Moreover, this procedure can be applied in other cross-validation contexts in which there is scarce information on mobility, by combining the available data with digital traces from social media.

\section{Results}

We processed the traces left by $40,000$ Flickr users in the U.S. (about 18 million pictures) in order to obtain mobility flow matrices which represent human flows between pairs of geographical nodes, looking for the collective mobility patterns of two types of human activities at different resolution scales: air travel and daily commuting. 
We also used two real mobility datasets as a ground truth: the RITA dataset of air travel in the U.S.~\cite{rita}, and the commuting data provided by the U.S. Census Bureau~\cite{census}. 
At each resolution scale, flows were aggregated into geographic basins. For air travel, our geographic basins were defined by the presence of an airport, while for the commuting network each basin corresponded to a U.S.county (see the {\em Methods} for more details on the ground truth mobility datasets). 

The construction of the Flickr flow matrices at the airport and county level from the users' traces is described in {\em Methods: Flickr-based flow matrices}. 
In short, we added a connection between pairs of basins, $i$ and $j$, every time a user took a picture in basin $i$ and the subsequent one in basin $j$. Each connection is then weighted by the total number of users who travelled between $i$ and $j$.  

The distance between the locations of two consecutive pictures taken by the same user ranges from a few meters to thousands of kilometers, showing a heavy-tailed behavior with an exponential cutoff (see the {\em Supplementary material} for more details). 
Thus, it is clear that users' traces can provide information about very different types of human displacements, ranging from a short walk within a city to long distance trips or international travel. 
In each case, and when modelling a particular type of mobility, it is important to consider only the relevant traces for that type of movements, otherwise, the effects of different activities will be mixed. 
We analyzed the effect of distance on both the daily commuting and the air travel, and we set a maximum trip distance of 100~km for daily commuting in the U.S. and a minimum trip distance of 500~km for U.S. air travel. The analysis is described in {\em Methods: Distance thresholds}.

After aggregating the trips by basin and filtering by distance, we obtain the Flickr flow matrices denoted as $\mathbf{F}_r=(f^r_{ij})$ for the air travel network, and $\mathbf{F_c}=(f^c_{ij})$ for the commuting network. 
Here, $f^c_{ij}$ and $f^r_{ij}$ represent the number of Flickr users travelling from basin $i$ to basin $j$ in each of the networks.

{\bf Model definition and fitting.} A model of human mobility should be able to predict real mobility flows. The aim of our work was to predict mobility flows of the air travel network and the U.S. commuting network, denoted as $\mathbf{Y_r}=(r_{ij})$ and $\mathbf{Y_c}=(c_{ij})$ respectively. 
Here, $r_{ij}$ represents the number of travellers between two airports $i$ and $j$, while $c_{ij}$ represents the daily amount of commuters between any two counties $i$ and $j$. 
Adopting a machine learning approach, we represent this prediction problem as a regression task in which the model is first trained, and then it is used to estimate the real flows, $\mathbf{Y_r}$ or $\mathbf{Y_c}$, which are the so called target values.

The classical gravity model (described in the {\em Methods}) estimates the target values as directly proportional to some powers of the population of the origin and destination basins and inversely proportional to some increasing function of the distance between them (typically, a power-law or exponential function). 
We will indicate this model as $\mathbf{G}(\alpha, \beta,\gamma; \mathcal{P})$, where $\alpha, \beta$ and $\gamma$ are the three exponents in the gravity law, and $\mathcal{P}=(p_i)$ is a vector containing the populations of the basins.

Analogously, travel flows estimated from Flickr traces can be described by a model in which we assume that the value $y_{ij}$, representing the mobility flow between two basins $i$ and $j$, is proportional to the number of users' trips from $i$ to $j$. 
We represent this model as $C_F\cdot\mathbf{F}$, where $\mathbf{F}$ represents the Flickr flow matrix at the corresponding resolution level (i.e., $\mathbf{F_r}$ or $\mathbf{F_c}$), and $C_F$ is a constant.

Our approach is to combine the flows of human mobility estimated from the two models under a stacked regression procedure~\cite{breiman1996stacked} in which each model is fitted alone, and then a linear regression determines the weight of each of them. 
In this way we combine the Flickr traces and the gravity model to improve the prediction of the real mobility patterns.

The incorporation of the traces will be defined by a linear function (we omit the subscript indices to generalize to any of the two resolution levels):
\[
\mathbf{H}(\alpha, \beta,\gamma,A,B; \mathcal{P}) = A\cdot\mathbf{G}(\alpha, \beta,\gamma; \mathcal{P}) + B\cdot\mathbf{F},
\]
where $\mathbf{H}=(h_{ij})$ represents our hybrid model, $\mathbf{G}$ is the gravity model and $\mathbf{F}$ is the Flickr flows matrix; $\alpha, \beta,\gamma,A,B$ are real-valued fitting parameters. We will fit the model by minimizing the following loss function $L$:
\[
L = \parallel\mathbf{Y} - \mathbf{H}(\alpha, \beta,\gamma,A,B; \mathcal{P})\parallel,
\]
where $\parallel\cdot\parallel$ denotes the Frobenius norm. Again, we removed the subscript indices, so that this equation represents what is done at each resolution level. The minimization process is made under the stacked regression assumption that each of the two component models can be fitted alone, and then a linear regression determines the weight of each of them.
It is important to notice that, as $\mathbf{Y_r}$ and $\mathbf{Y_c}$ are sparse matrices, we only evaluate $L$ for those connections for which there is a positive flow of travellers.

{\bf Model evaluation.}
The performance of a learning model is measured by its capacity to generalize to flows that were not known during the training step. 
Then, in order to validate our model we used a cross-validation strategy in which we fitted the model using only part of the target flows $\mathbf{Y}$ (training set) and then we tested its performance using the remaining flows (test set). 
We measured the performance in terms of Pearson correlation coefficient $\rho$ and the determination coefficient $r^2$ between the target flows and the predicted flows.

Our procedure follows a $10$-fold cross-validation scheme~\cite{hastie2009elements}: the real dataset is divided into $10$ parts or folds, and for each fold we use the $10\%$ of the target values as test set, and the remaining $90\%$ as training set. Thus, each sample in the dataset was tested once, using a model that was not fitted with that sample.
The performance of the different models is shown in Table~\ref{table_performance} in terms of the Pearson correlation coefficient $\rho$ and the determination coefficient $r^2$ between the real flows and the predicted flows.
The table shows also the results for the gravity model alone ($A=1$, $B=0$) and the Flickr model alone ($A=0$, $B=1$). To note, all the results were cross-validated, i.e. each individual flow was estimated with a fit that excluded it from the training set.

\begin{figure*}[!htb]
 \centering
\includegraphics[width=16cm]{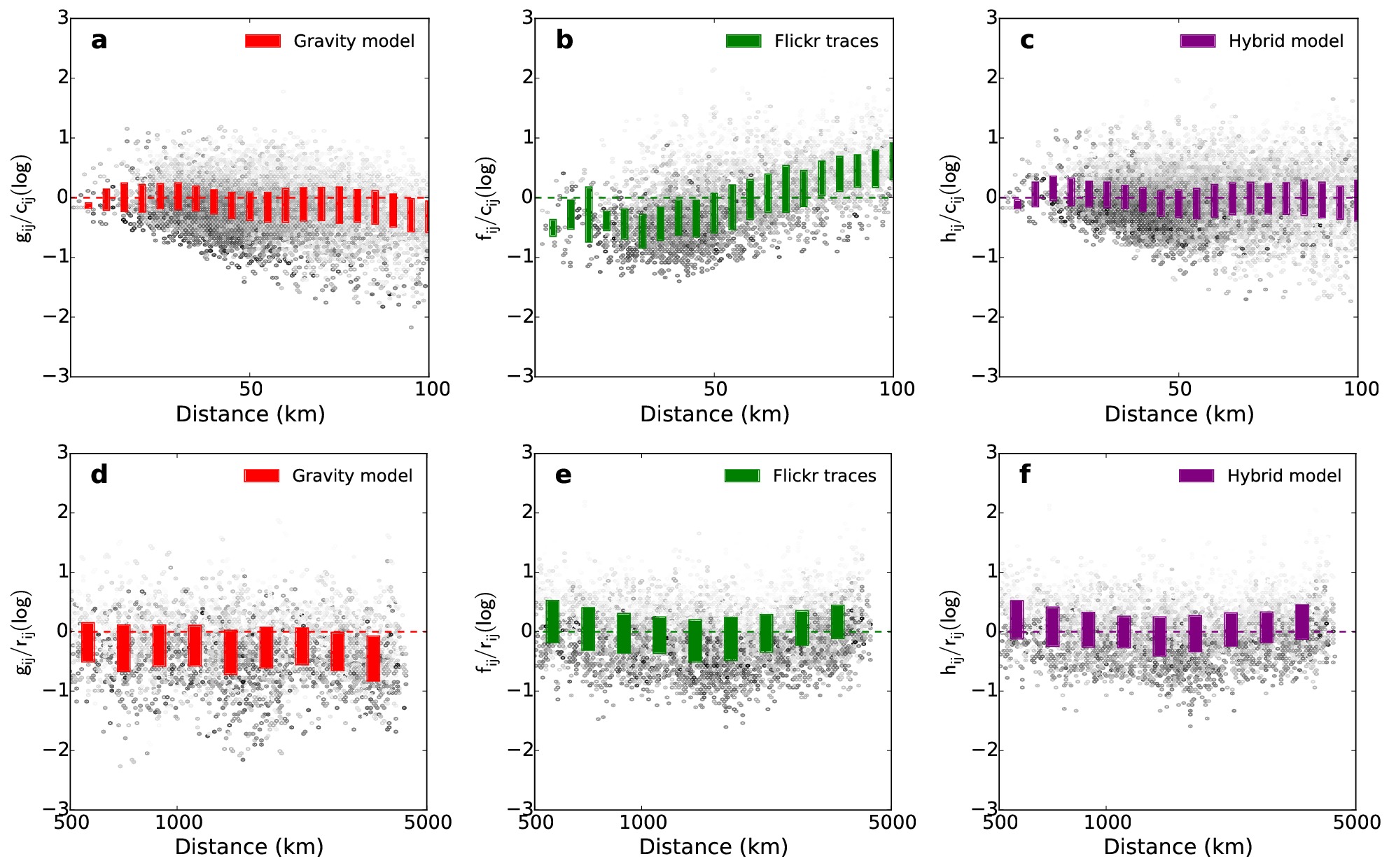}
  \caption{\label{figure_ratios}{{\bf Prediction ratio for large flows as a function of distance.}  In each panel, grey dots correspond to the flows between basins as predicted by the gravity model ($g_{ij}$), by the Flickr traces ($f_{ij}$), and by the hybrid model combining both ($h_{ij}$), in relation to the real flows, and as a function of distance between basins. Flows predictions were all made under a 10-fold cross-validation scheme. {\bf (a, b, c)} Commuting network of the U.S. {\bf (d, e, f)} Air-transportation network of the U.S. (We only show flows above $100$ passengers). The color intensity at each point represents the total passenger flows aggregated under a certain distance and prediction ratio. Boxplots represent interquartile ranges.}}
\end{figure*}

Notice that the low determination coefficients for the gravity model alone are related to its limitations for capturing the large flows across distant cities, due to the fast decay of the model as a function of distance. This had also been observed in~\cite{simini2012universal}. 
Indeed, by plotting the prediction ratio of the gravity model for flows above $100$ passengers in the red curves of Fig.~\ref{figure_ratios} we clearly observe that the gravity model tends to underestimate large flows. 
Instead, the hybrid model captures correctly the large distance flows (violet curves of Fig.~\ref{figure_ratios}).

\begin{table*}[!htb]
	\centering
      \begin{tabular}{ccccc}
	\toprule
	{\bf Model} &  \multicolumn{2}{c}{{\bf Commuting}} &  \multicolumn{2}{c}{{\bf Air travel}}\\
	 & $\rho$  & $r^2$ & $\rho$  & $r^2$ \\
	\midrule
        Gravity alone & 0.69 & 0.41 & 0.68 & 0.40 \\
        Flickr alone & 0.69 & 0.47 & 0.78 & 0.62 \\
        Hybrid model & 0.79 & 0.62 & 0.84 & 0.72 \\
	\bottomrule
      \end{tabular}
\caption{\label{table_performance}{\bf Cross-validated model performance ($10$-fold cross-validation).} The table shows the performance of our hybrid model in terms of the Pearson correlation coefficient $\rho$ and the determination coefficient $r^2$. We also show the results for the gravity model alone and the Flickr traces on their own. All the values were produced under a $10$-fold cross-validation scheme.}
\end{table*}

\begin{figure*}[!htb]
 \centering
\includegraphics[width=17cm]{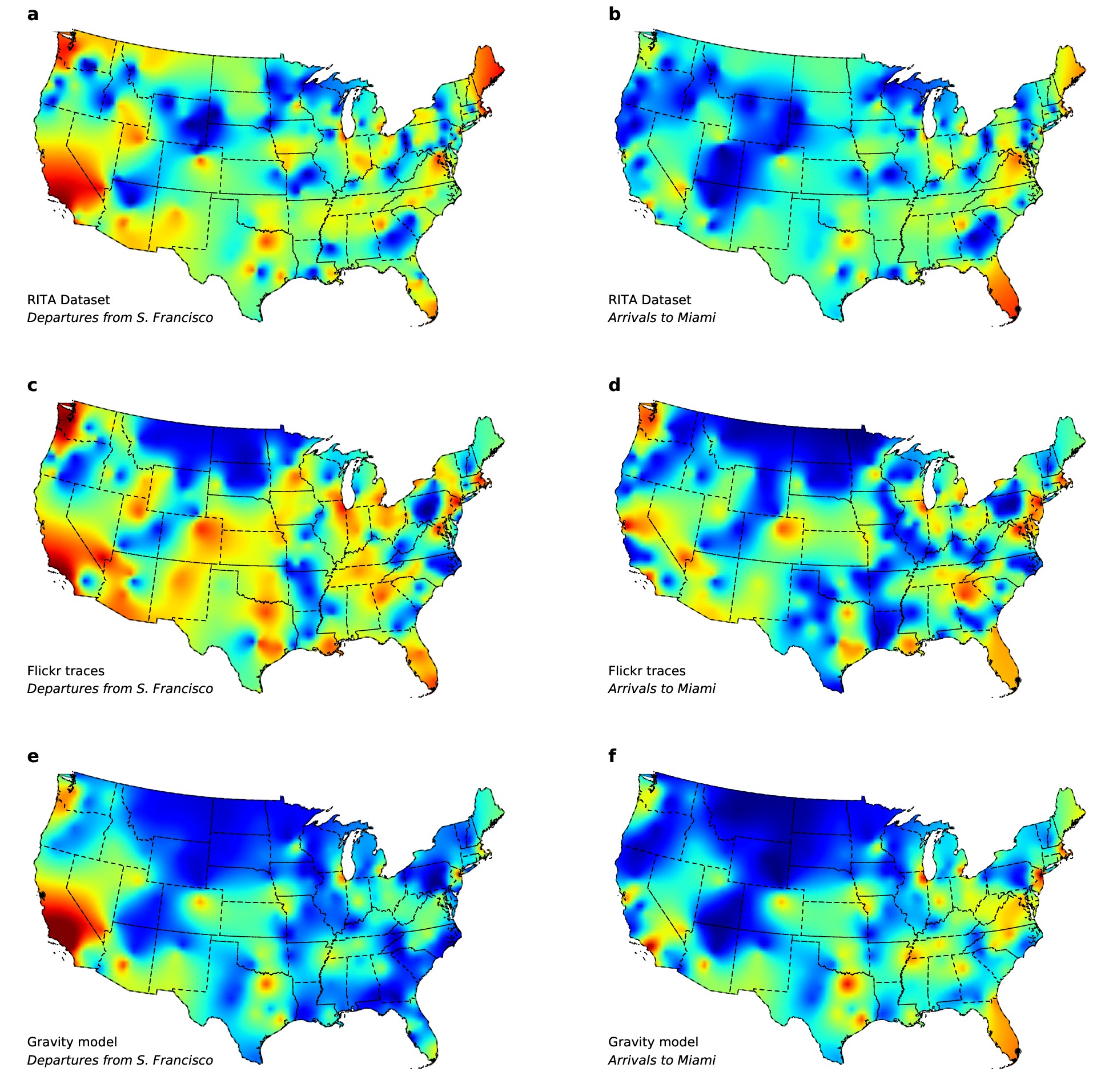}	
  \caption{{\bf Departures and arrivals in the air transportation network.} Heatmaps representing the distribution of predicted trips departing from San Francisco towards any point in the U.S. (left column), or arriving to Miami from any point in the U.S. (right column).  {\bf (a, b)} Ground-truth from the RITA air travel dataset.  {\bf (c, d)}  Flickr traces. {\bf (e, f)} Gravity model.}\label{fig_heats}
\end{figure*}

On the other hand, according to Table~\ref{table_performance}, the performance of the Flickr model alone is comparable to that of the gravity model for the commuting network, and is significantly higher in the air travel network. 
This reveals that Flickr users' trips can be a good proxy of collective human mobility at different resolution scales, and can be particularly useful when real data for fitting the gravity model is not available. 

\begin{figure*}[!htb]
 \centering
\includegraphics[width=16cm]{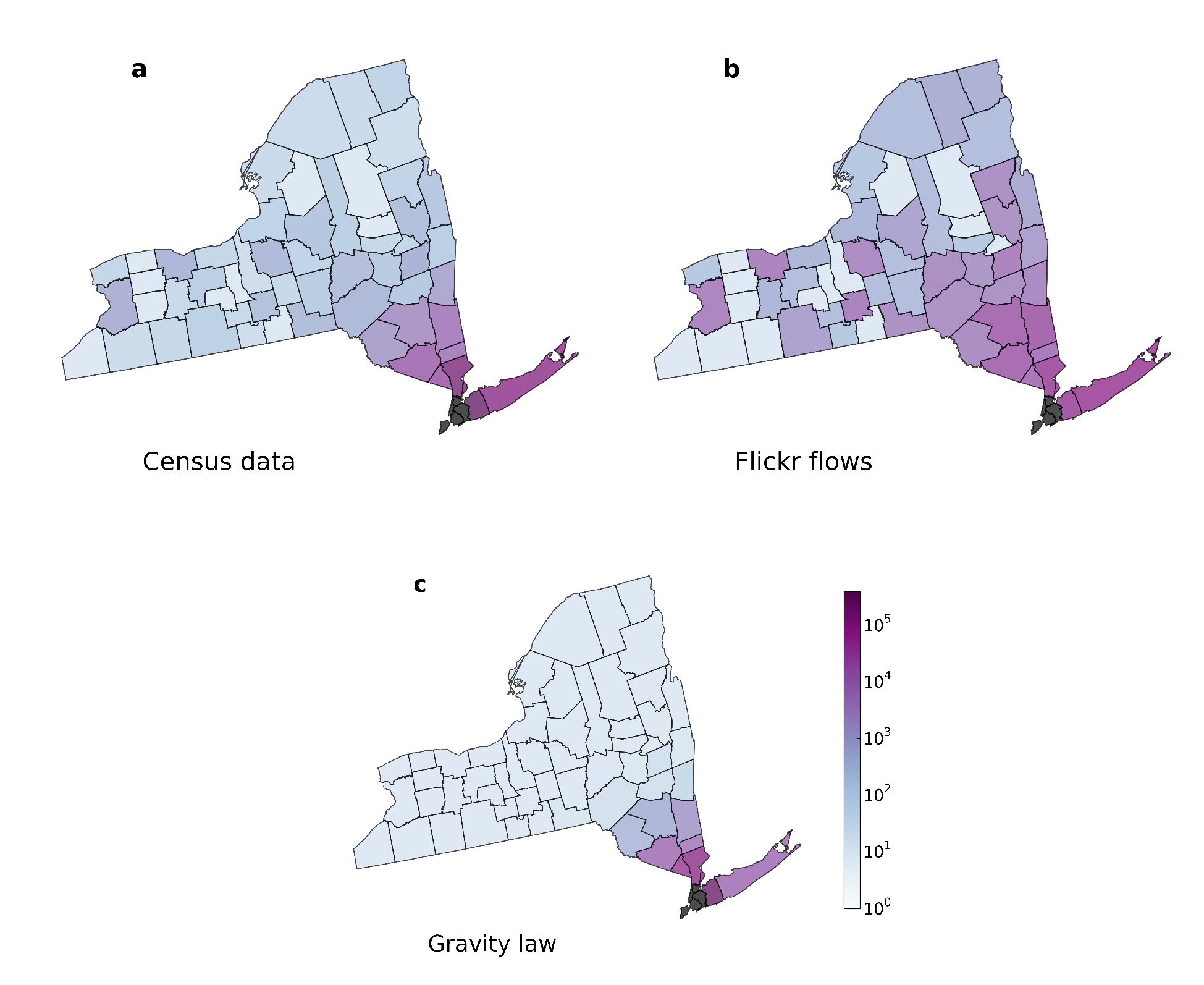}
  \caption{{\bf Daily commuting to New York City.} Each panel shows the daily amount of commuters arriving to New York City from the different counties of the New York State. Here, we aggregated all the commuters traveling to the 5 boroughs of New York City. {\bf (a)} U.S. Census Bureau data. {\bf (b)}  Flickr traces. {\bf (c)} Gravity law. Flickr traces are shown without taking into account any distance treshold.}\label{fig_comms_new_york}
\end{figure*}

A more qualitative evaluation of the geographical spanning of Flickr traces is offered in Fig. ~\ref{fig_comms_new_york} and~\ref{fig_heats}, which show the distribution of the origins and destinations of travellers according to the real datasets, the Flickr traces and the gravity model. 
Fig.~\ref{fig_comms_new_york} shows the commuting patterns in the New York State, where the color intensity represents the amount of people commuting from one county to New York City (in black).
Here, we observe that the gravity law correctly captures the flows from neighbouring counties, which in fact represent more than $95\%$ of the commuting flows, but it underestimates long distance flows. 
Instead, the traces from Flickr have a slower distance decay, more in accordance with census data. 
Something similar is observed in the air travel network, as depicted in Fig.~\ref{fig_heats}. Looking at trips departing from San Francisco (left panels), the gravity law correctly predicts that the largest flows are those towards Los Angeles (CA), San Diego (CA), Las Vegas (NV) and Seattle (WA), but those directed to the East Coast are generally underestimated. 
For the arrivals to Miami (right panels) the gravity shows a good performance. In both cases, the flows from Flickr are a representative sample of all the territory. This large span is also confirmed when we observe the trip distance distribution of Flickr users (see the {\em Supplementary Information}).

The incorporation of the Flickr traces into the gravity model produced a significant increase in the predictive performance both for daily commuting as for air travel in the U.S., as shown in the last row of Table~\ref{table_performance}. 
We observed a relative increase of $\sim 8-20\%$ in the Pearson correlation $\rho$ between the model predicted flows and the real flows. 
The predictive power has also improved, as verified by a relative increase of $>50\%$ in the determination coefficient $r^2$ (i.e., to what extent the model accounts for the real flows). The ratio between predicted flows and real flows as a function of distance has also improved, as shown in the right panels of Fig.~\ref{figure_ratios}.

\begin{figure*}[!htb]
 \centering
\includegraphics[width=16cm]{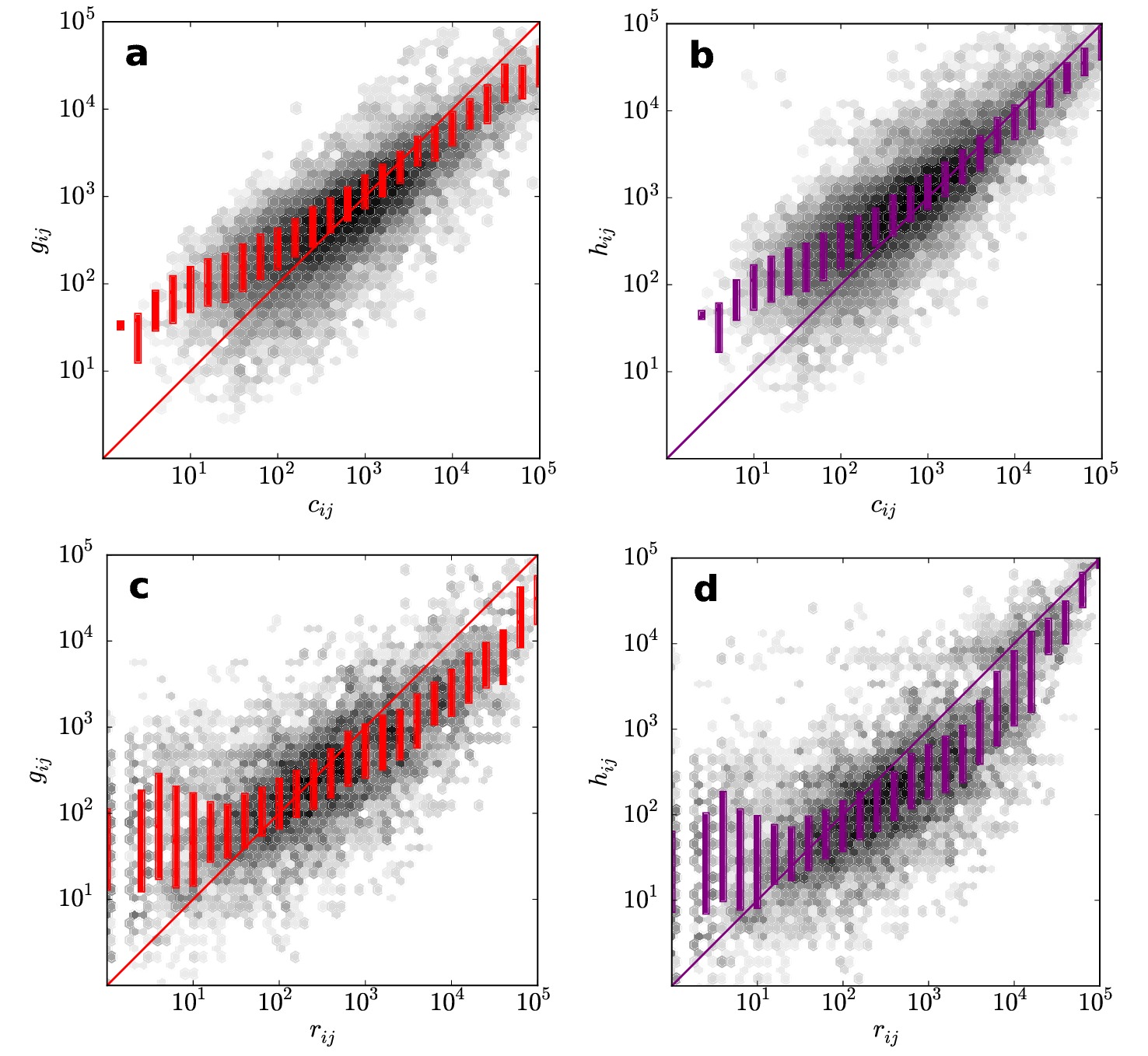}	
  \caption{{\bf Mobility models predictions} Each point in the 2D-histograms represents flows with some real/estimated flow value relation. The color of the points in a gray scale represents the frequency values. The boxplots in each panel correspond to the interquartile ranges. {\bf (a, b)} U.S. Commuting network. {\bf (c, d)} U.S. air travel network.}\label{fig_scatter}
\end{figure*}

\begin{figure*}[!htb]
 \centering
\includegraphics[width=16cm]{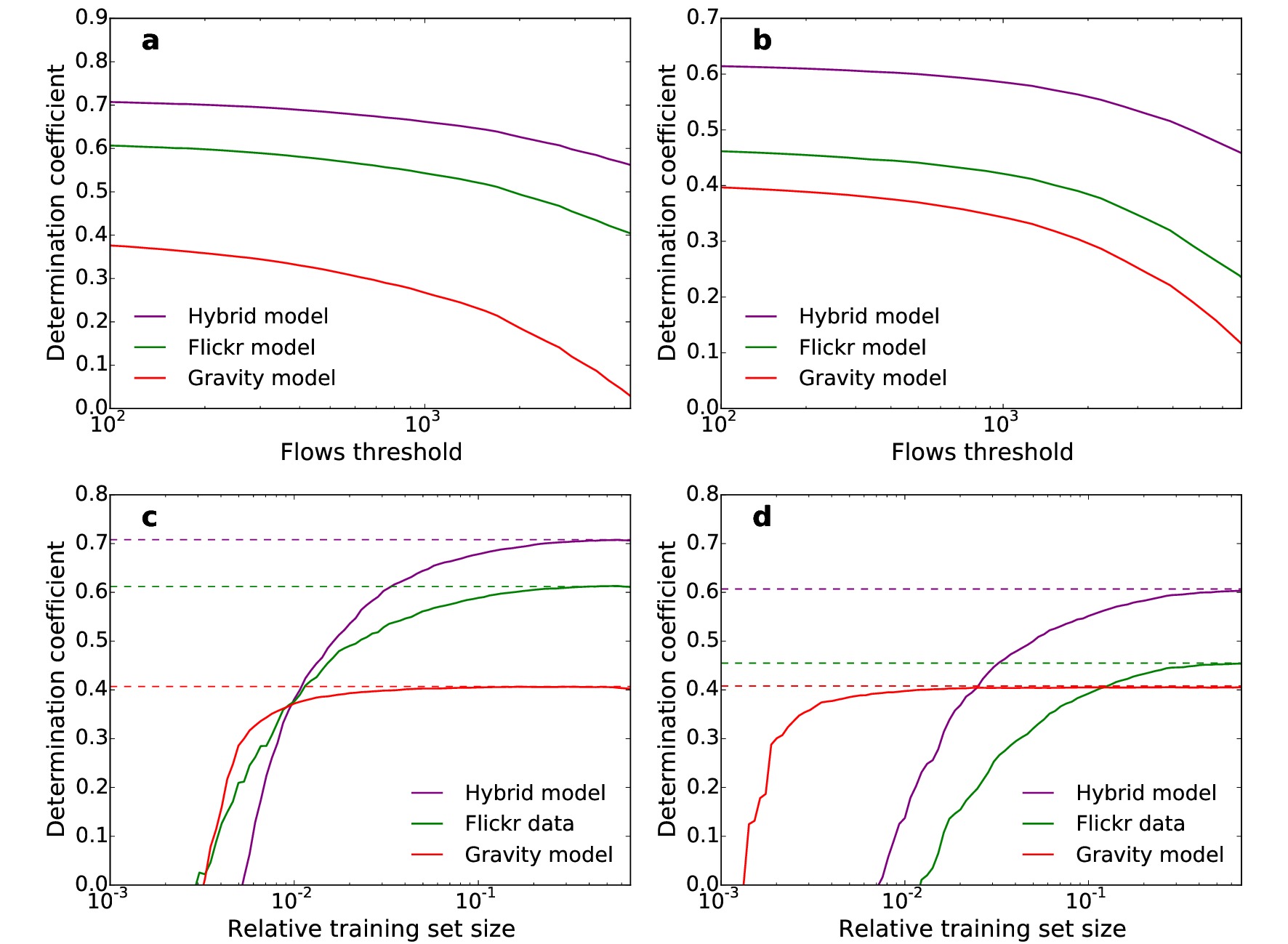}	
  \caption{{\bf (a, b)} {\bf Determination coefficient for the prediction of large flows, as a function of the flow threshold.} Performance of the models measured by the determination coefficient $r^2$ when restricted to pairs of nodes for which the real flows are above a given threshold value. {\bf (a)} Air transportation data. {\bf (b)} Commuting data. {\bf (c, d)} {\bf Prediction error rate for different training set sizes.} Prediction error in terms of the determination coefficient, as a function of the training set relative size. The training set sizes were varied in a logarithmic scale. Dotted lines correspond to the maximum attained values. {\bf (c)} Air transportation data. {\bf (d)} Commuting data.}\label{fig_large_flows}
\end{figure*}

Fig.~\ref{fig_scatter} shows several 2D-histograms for the gravity model alone and for the hybrid model, together with boxplots grouped by the real flow values. The plots compare the model predictions to the real mobility data for commuting and air travel, and each cell represents pairs of nodes (counties in the case of commuting, and airport basins for air travel). 
Despite the fact that most of the interquartile ranges do not change in average, it is interesting to observe the changes on the right side of the pictures: for the models incorporating Flickr data, the interquartile ranges become shorter for large values of flows, and at the same time they are better aligned with the diagonal line representing the perfect matching between model and real data. 
These means that the improved model is both more precise and more exact for pairs of nodes connected by large mobility flows.
This behavior is clearly seen if we compute the determination coefficient of the model after filtering flows above a certain value. 
We show this analysis in Fig.~\ref{fig_large_flows} (panels a,b), where the $x$-axis represents the flow threshold and the $y$-axis is the determination coefficient between the models and the real flows, restricted to pairs of nodes for which the real flow is larger than the threshold. The contribution of the Flickr traces is particularly significant in the air-travel network, where they duplicate the gravity model in a predictive capacity. In the commuting network both models are close, but the hybrid model almost duplicates the gravity performance for flows above $2000$ travellers.

In the following two subsections we validate the model under different geographical and random data availability constraints. The {\em Supplementary Information} also includes a test based on the Sorensen-Dice coefficient, in which we evaluated the model performance for different distances and populations. This test shows that flows towards largely populated basins are the ones most improved by the hybrid model.

{\bf Prediction under data availability constraints.} We evaluated the power of our regression model as a function of the training set size in order to see the minimum amount of data required to reach the largest improvement in the predictive power of both the gravity law and the Flickr flows. 
The analysis was performed under a cross-validation scheme with repeated random subsampling (bootstrapping). The advantage of this procedure is that the training set size can be varied in small steps (while $k$-fold cross-validation only allows for a minimum training set size of $0.5$). The results of this test are shown in Fig.~\ref{fig_large_flows}. For the air transportation network, we observe that with a training size of $1\%$ the gravity law is already close to its best prediction levels, while from that value onwards the assimilation of flows from the Flickr traces starts producing an improvement in the performance. With a training size of $3\%$ this improvement is already quite significant. In the commuting network, training with $1\%$ of the ground-truth flows is also enough for the gravity law, but we need a $10\%$ of Flickr flows in order to observe a significant improvement in the model performance.

{\bf Spatial cross-validation: U.S. West Coast vs. U.S. East Coast.} To extend the analysis of the generalizability of the model, we performed a geographical $2$-fold cross-validation in which we split the contiguous United States into two roughly symmetric parts, taking by reference the meridian $-102^\circ$. 
We trained the model in one half of the U.S. and we then used it to predict the mobility flows inside the other half. 
To make our test independent from the specific choice of the spatial partition, we disregarded crossed flows, that is, the traffic flows connecting two points in different halves of the partition. 
Table~\ref{table_performance2} shows the performance in terms of the Pearson correlation $\rho$ and the determination coefficient $r^2$. We see that, in this case, the performance of Flickr traces is inferior with respect to the $10$-fold cross-validation values of Table~\ref{table_performance}, denoting the presence of spatial inhomogeneities in the users' activities across the U.S., especially for the commuting network. However, the performance of the hybrid model is still improved by the assimilation of the Flickr flows. 

\begin{table*}[!htb]
	\centering
      \begin{tabular}{ccccc}
	\toprule
	{\bf Model} &  \multicolumn{2}{c}{{\bf Commuting}} &  \multicolumn{2}{c}{{\bf Air travel}}\\
	 & $\rho$  & $r^2$ & $\rho$  & $r^2$ \\
	\midrule
        Gravity alone & 0.72 & 0.41 & 0.68 & 0.40 \\
        Flickr alone & 0.60 & 0.34 & 0.72 & 0.43 \\
        Hybrid model & 0.78 & 0.46 & 0.81 & 0.49 \\
	\bottomrule
      \end{tabular}
\caption{\label{table_performance2}{\bf Geographically cross-validated model performance (U.S. West Coast vs. U.S. East Coast).} Performance of our hybrid model in terms of the Pearson correlation coefficient $\rho$ and the determination coefficient $r^2$. The predicted values were produced by training the model in the West Coast and then validating in the East Coast and viceversa.}
\end{table*}

\section{Discussion}

Theoretical models of mobility offer the possibility of predicting human flows when calibrating data is available, which is the case in most developed countries, but not in all the world. The gravity model is currently the prevailing one, and has been successfully used in different contexts and scales. However, it has some limitations, as the underestimation of large flows and its fast decay with distance. On the other hand, the more recent radiation model requires less calibration data -- only the population distribution -- but has a limited geographical spanning, being usually applied for commuting at regional scale.

While in some cases first-principled theoretical approaches can be pursued \cite{ren2014predicting}, the increasing amount of geolocalized data publicly available through the Internet and social media suggests a different perspective for modelling mobility, which is the incorporation of large volumes of digital traces into theoretical models. 

In this work we followed this approach and we showed that geolocalized traces collected from social media as those from Flickr can successfully inform a predictive model of collective human mobility. 
Even though individual mobility trajectories can be noisy, their aggregation is representative of collective mobility patterns. 
Our results showed both qualitatively and quantitatively the high performance of Flickr traces in the predictions, at different distances and for different resolution levels. 

The use of traces from Flickr as a proxy for human mobility has also been explored by Barchiesi et al.~\cite{Barchiesi150046}, who successfully modelled the flows between the $20$ largest cities in the UK using traces from $16,000$ individuals, and considering a threshold value to fix the cities' diameters. They showed the agreement of their predictions with travel data obtained from surveys. Our results strengthen the capabilities of digital traces for predicting human mobility, as we show that they can be assimilated into the gravity to produce a data-driven model. 
Our model was tested at different resolution levels, for flows among over $200$ nodes for the U.S. air travel network, and over $3,000$ counties for the commuting network. Our predicted flows span over cities and counties of variable area and population, and at quite different resolution scales. 

The assimilation of Flickr traces to produce a data-driven model addressed the limitations of the gravity model, as the availability of data and the underestimation of some large distance flows~\cite{simini2012universal, masucci2013}. Moreover, considering the potential data availability limitations in many situations, our cross-validation with bootstrapping approach showed that a small sample of data ($\sim 1\%$) is enough for its calibration. Thus, by assimilating the traces from Flickr users into the classical gravity model, we improved its predictions in several ways: we handled the underestimation issues, we accounted for missing calibration data and we reinforced the model by providing it with individual traces.

Finally, we note that we used an open dataset as is the Flickr 100M dataset~\cite{thomee2015yfcc100m}, which can be publicly accessed by anyone. However, Flickr is far from being the only appropriate platform for modelling mobility, and our research suggests the possibility of assimilating data from several sources simultaneously, in order to capture other types of activities and users. A future line of research also involves exploring other assimilation techniques different from the simple stacked regression.

In conclusion, our results expose two new directions in the modelling of mobility flows: first, that the assimilation of digital traces can reinforce classical mobility models as the gravity improving its predictions significantly. Second, our cross-correlation study proved that with a small number of real flows our model can be adjusted to give predictions close to its optima. These contributions are in line with current challenges in the study of spreading processes and social systems, as the assimilation of geolocalized data into network models and the construction of models at appropriate resolution scales.

\section{Methods}

\subsection{\bf Dataset description}

We use two large real datasets as ground-truth for human mobility at quite different resolution levels:
\begin{itemize}
\item {\bf Air travel in the U.S.} The RITA dataset from the {\em Airline Origin and Destination Survey}~\cite{rita} collected by the {\em U.S. Bureau of Transport Statistics}, contains a $10\%$ sample of all the domestic itineraries in the U.S. We used the subset correspondent to 2014, which comprises $\sim$ 14 million air tickets between $466$ airports.
\item {\bf Commuting in the U.S.} This dataset obtained from the {\em U.S. Census Bureau}~\cite{census} contains data on $\sim$ 36 million commuters from the period $2009-2013$ at the county level. It specifies a {\em ``home county''} and a {\em ``working county''} for each commuter.
\end{itemize}

\subsection{\bf Ground-truth flow matrices}

The ground-truth flow network for the air transportation in the U.S. was built using the RITA dataset~\cite{rita}. We define an airport basin as the area covered by an airport (i.e., the set of points for which that airport is the closest one). Thus, the partition of the U.S. territory into basins is built as the Voronoi tessellation given by the airport coordinates. We note that when two airports are at less than 30~km of distance, we consider that they represent a single airport basin (because they serve the same metropolitan area) and we replace them by a single airport before computing the Voronoi tessellation. We also generalized this idea to connected components of airports at less than 30~km.

Each ticket in the RITA dataset contains an itinerary formed by several {\em (coupons)}. Each coupon contains information about the origin and destination airport of the trip and the number of passengers, and it also points out whether the destination was a stopover or either the passengers remained there. Those destinations in which the passengers did other than just a stopover are marked as {\em trip breaks} (TB); the first and last airports from an air ticket are always TB's. By removing the stopovers from the itinerary we manage to clean the flow network from the presence of airport hubs, which do not represent the real passengers destination. For each itinerary, we obtain a list of destinations $(d_1, d_2, ..., d_n)$ ($d_1$ is the departing city), and we use it to build the flow network between the airport basins.

For the commuting dataset~\cite{census}, each sample contains the {\em ``home county''} and {\em ``working county''} of one worker. We consider the list of U.S. counties as nodes for the commuting flow network, and for each pair of counties $(c_i, c_j)$ we put a weighted link counting the number of workers that live in one of them and work in the other.

We use the notation $\mathbf{Y_r}=(r_{ij})$ and $\mathbf{Y_c}=(c_{ij})$ for the adjacency matrices which describe the airport and commuting ground-truth flows, respectively. We put the diagonal elements of both matrices to zero, because we shall not consider users that take a flight inside the same airport basin (we only observed $5$ cases) or commuters which do not change county from home to work.

The air travel matrix $\mathbf{Y_r}$ contains $204$ airport basins with $30,472$ links, with an amount of $\sum_{ij}{r_{ij}}=40$ million flows. The commuting matrix $\mathbf{Y_c}$ covers the $3,144$ U.S. counties and has $55,578$ links, describing the pattern of about $74$ million commuters.

\subsection{\bf Flickr-based flow matrices}

The traces left by Flickr users when they take and upload pictures are given by the coordinates of their geotagged pictures, ordered by the time in which they were taken. We only consider timestamped, geolocalized pictures taken in the U.S. For each user we obtain an array of pictures $(p_1, p_2, ..., p_n)$ sorted by timestamp. At a particular resolution level (airport or county) we will consider that the user makes a trip when two consecutive pictures have coordinates belonging to different basins. Then, we aggregate all the users' trips into a Flickr flow matrix.
\begin{itemize}
\item{{\bf County-level Flickr flow matrix:} We assign each picture to a county by considering the county borders as defined in the MAF/TIGER geographic database of the U.S. Census Bureau. We count one flow between counties $(i,j)$ when two consecutive pictures $(p_i, p_{i+1})$ are taken in counties $i$ and $j$ respectively. We do not consider successive pictures in which the user does not change county.}
\item{{\bf Airport-level Flickr flow matrix:} We choose the airport basin closest to each picture. We count one flow between two airport basins $(i,j)$ whenever two successive pictures are taken in basins $i$ and $j$ respectively. We shall not consider successive pictures in which the user does not change airport basin.}
\end{itemize}

We note the adjacency matrices of these networks as $\mathbf{F_c}=(f^c_{ij})$ and $\mathbf{F_r}=(f^r_{ij})$. Both of them have zero diagonals.
In total, we observed $\sum_{ij}{f^r_{ij}}\approx350,000$ trips between airport basins and $\sum_{ij}{f^c_{ij}}\approx520,000$ trips between counties. The flow networks contain $\approx26,000$ and $\approx150,000$ nonzero elements, respectively.

\subsection{\bf Distance thresholds}

\begin{figure*}[!htb]
 \centering
 \includegraphics[width=16cm]{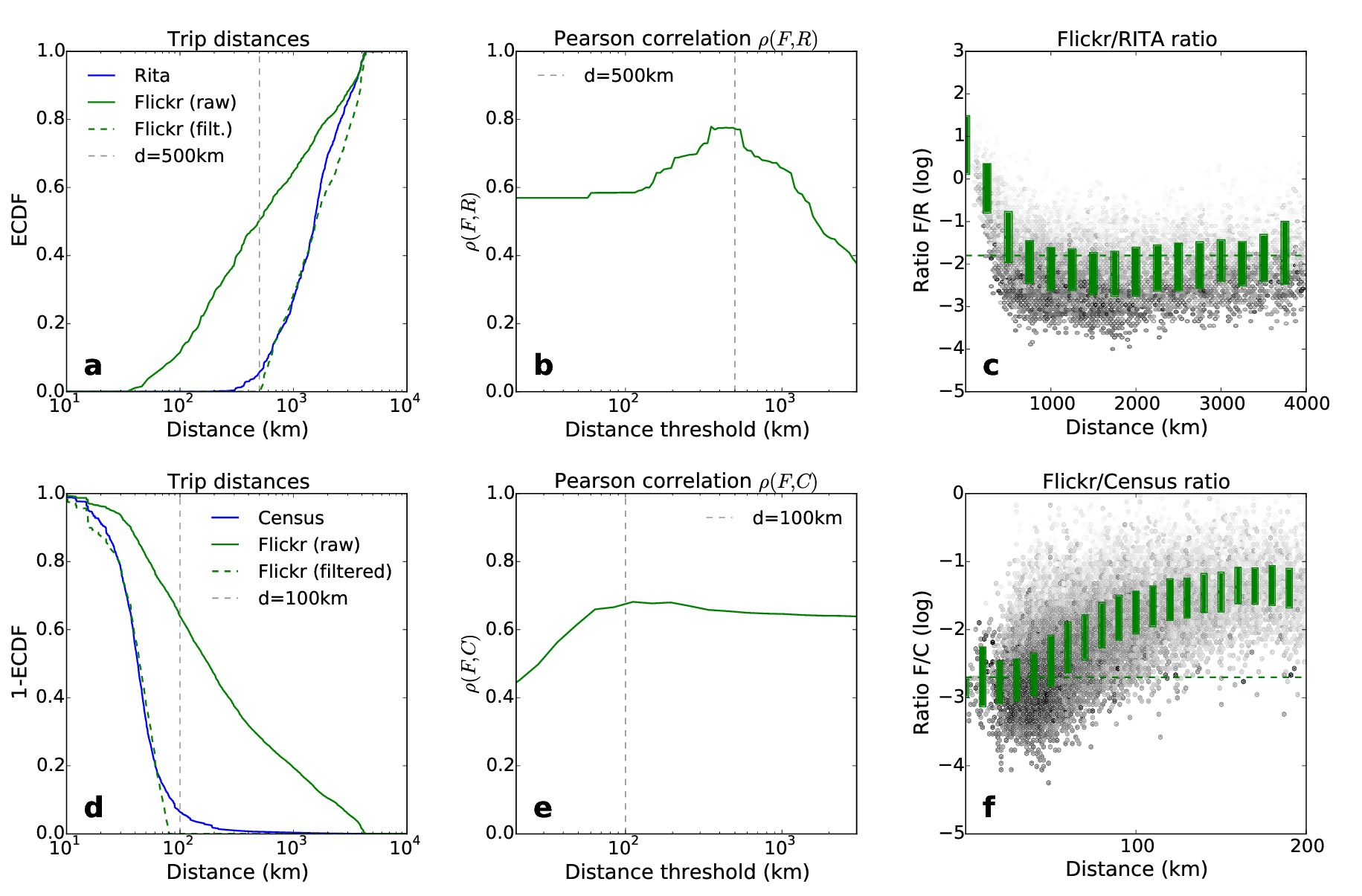}	
  \caption{{\bf Distance calibration for the U.S. air transportation network and the U.S. commuting network.} A correct fitting of the human mobility networks based on geolocalized traces requires aggregating the latter at the appropriate resolution level (basins or counties, respectively) and filtering those ones associated with the correspondent mobility type. Distance is an important calibrator for this task, as can be seen from: {\bf (a,d)} Cumulative distribution of the trip distances in Flickr compared with the ground-truth (the RITA dataset in the case of air travel, and the census information for commuting); {\bf (b,e)} Pearson correlation between Flickr flows and the ground-truth flows when considering Flickr trips above a distance threshold; {\bf (c,f)} Flow ratio between Flickr and the ground-truth flows for a pair of (source, destination) nodes --basins or counties-- as a funtion of the distance between them. (The green lines represent the linear regression coefficient between Flickr and the ground-truth flows. {\bf (a,b,c)} Air transportation network; {\bf (d,e,f)} Commuting network.}\label{figure_distance}
\end{figure*}

As the activity of Flickr users involves different modalities of mobility, it has to be correctly filtered before comparing it against the ground-truth flows. We observed that an important variable for this task is the distance between nodes. In Fig.~\ref{figure_distance} we compare the ground-truth flows for air travel and commuting against the Flickr flows at the county level and the airport level respectively. We observe that the Flickr flows have good agreement with the ground-truth when we consider distances above 500~km for air travel, and below 100~km for commuting. In fact, if we remove those flows from the data, we observe that the Flickr users trip distance distributions are consistent with the ground-truth trip distributions. The threshold distances are also revealed as the value for which the correlation between the real flows and the Flickr predicted flows is maximum.

\subsection{Gravity model}

The gravity model considers that the flow between two nodes $(i,j)$ is directly proportional to some power of their populations and inversely proportional to an increasing function of the distance between them:
\[
g_{ij}=K\cdot\frac{P_{i}^\alpha\cdot P_{j}^\gamma}{d(i,j)^\beta}.
\]
We adjusted the gravity model using a linear regression in the logarithmic scale and following the approach of Balcan et al.~\cite{balcan2009multiscale}: we chose a power law of the distance $f(d)=d^{-\beta}$ for the air travel network and an exponential decay $f(d)=e^{-\beta\cdot d}$ for the commuting network, which provided the best results. 
The population information for the fitting was extracted from the public GeoNames database~\cite{geonames_url}, and the population of a basin was computed as the sum of the populations of all cities inside that basin.

\section*{Acknowledgements}	
This work has been partially funded by the EC FET-Proactive Project MULTIPLEX (Grant No. 317532) to M.T. and C.C. The authors also acknowledge support from the ``Lagrange Project'' of the ISI Foundation funded by the Fondazione CRT and from the ``S3 Project'' funded by the Compagnia di San Paolo.

\section*{Supplementary Information}

\subsection*{Distribution of Flickr users' trip distances}

\begin{figure*}[!htb]
 \centering
\includegraphics[width=10cm]{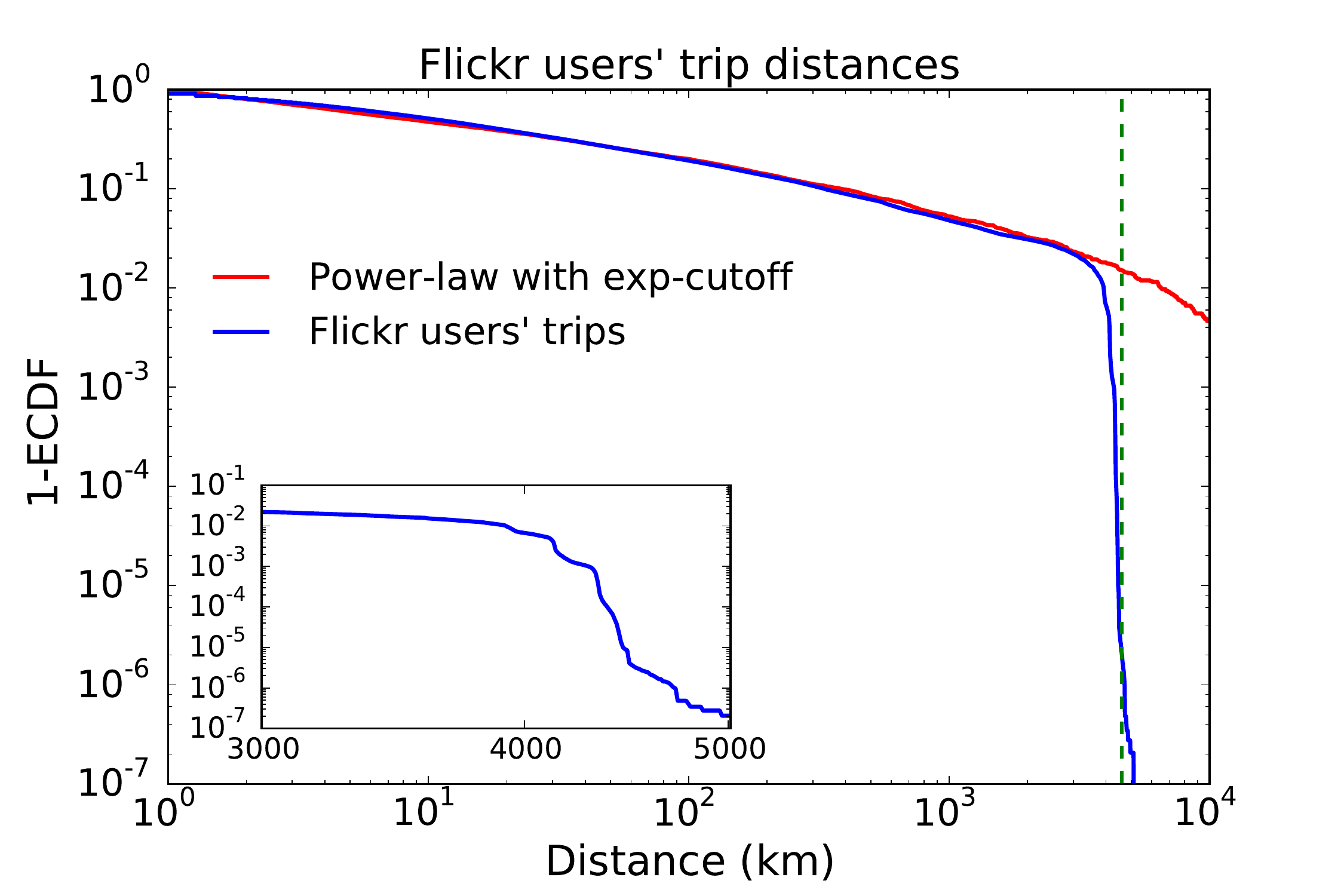}	
  \caption{{\bf Trip distances.} The trip distances of Flickr users follow a power-law distribution with exponential cut-off, bounded by the extension of the continental U.S., whose diameter (maximum distance between two points) is about $4,000$ kilometers. This distribution is in accordance with previous research on individual human mobility~\cite{gonzalez2008}.}\label{fig_trip_distances}
\end{figure*}

Figure~\ref{fig_trip_distances} plots the trip distances of the Flickr users in the U.S., showing a power-law distribution with exponential cut-off.

\subsection*{Large flows between airports predicted by Flickr}

\begin{figure*}[!htb]
 \centering
 \includegraphics[width=8.5cm]{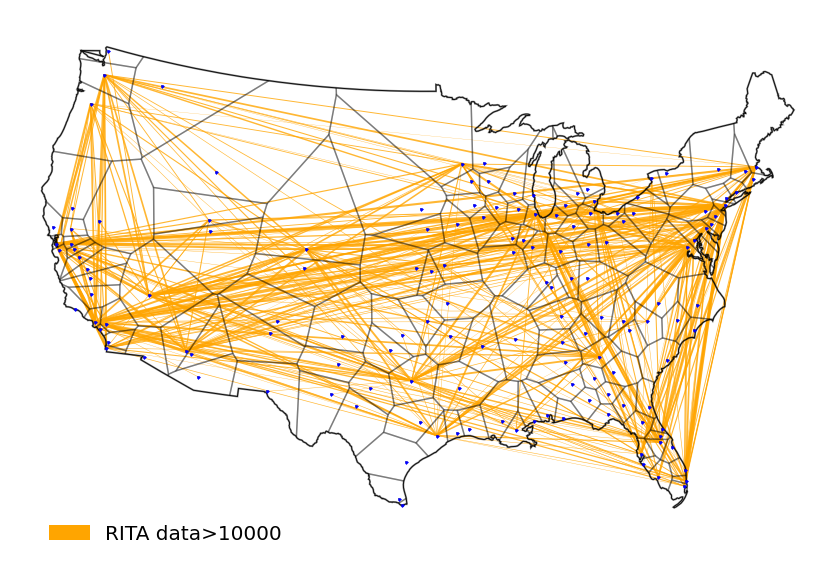}	
 \includegraphics[width=8.5cm]{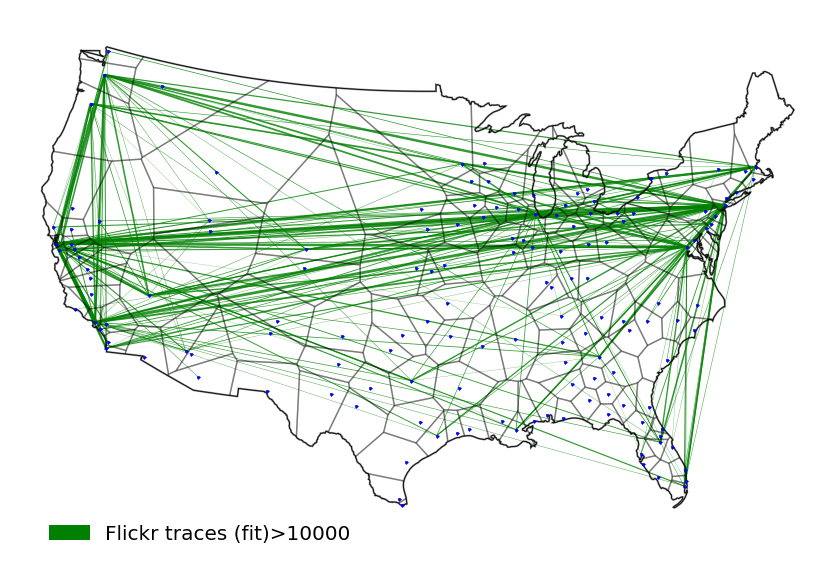}	
  \caption{{\bf Large air transportation flows captured by Flickr.} {\bf a} Air flight connections with more than $10,000$ according to the RITA dataset. {\bf b} Flickr flows with more than $10,000$ trips (after linear re-scaling).}\label{figure_maps}
\end{figure*}

Figure~\ref{figure_maps} shows that the Flickr trips are highly correlated with the air travel dataset when observing the pairs of airport basins with large flows of passengers. The green flows represent the connections predicted to have more than $10,000$ passengers by Flickr, after a simple linear re-scaling with the real data.

\subsection*{Large commuting flows predicted by Flickr}

\begin{figure*}[!htb]
 \centering
 \includegraphics[width=8.5cm]{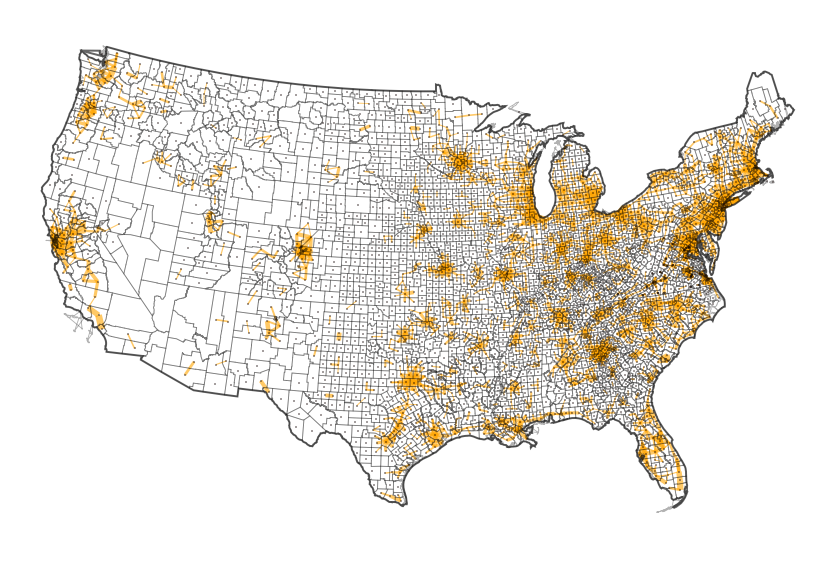}	
 \includegraphics[width=8.5cm]{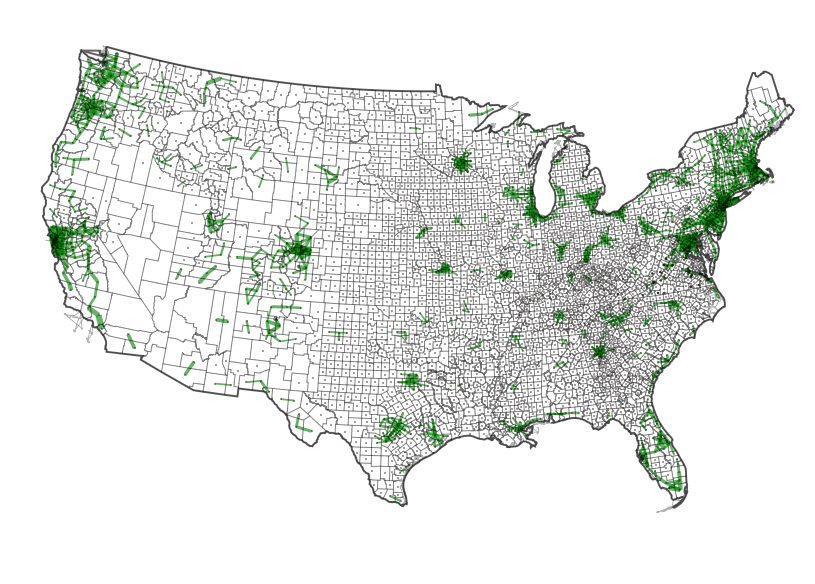}	
  \caption{{\bf Large commuting flows captured by Flickr.} {\bf a} Commuting flows with more than $1,000$ according to the U.S. Census dataset. {\bf b} Flickr flows with more than $1,000$ trips (after linear re-scaling).}\label{figure_maps2}
\end{figure*}

Figure~\ref{figure_maps2} shows that the Flickr trips are highly correlated with the commuting census data when observing the pairs of counties with large flows of commuters. The green flows represent the connections predicted to have more than $1,000$ passengers by Flickr, after a linear re-scaling with the real data.

\subsection*{A Sorensen-Dice coefficient based test}

In the main text we thoroughly explored the performance of the hybrid model under geographical and random sampling {\em (bootstrapping)} constraints. Here we will analyze the performance for different flows subsets organized by distance and destination population, using a test based on the Sorensen-Dice similarity coefficient. The Sorensen-Dice similarity between two sets $A$ and $B$ is defined as
\[
s(A,B)=\frac{2\cdot|A\cap B|}{|A|+|B|} \enspace .
\]
Here we follow the modification introduced in~\cite{lenormand_2012} (named as {\em ``Common part of commuters'' or CPC}), and also used by~\cite{masucci2013, yang2014limits, subsah_2015}, for comparing the predicted flows $\textbf{H}=(h_{ij})$ against the real flows $\textbf{Y}=(y_{ij})$:
\[
CPC(\textbf{H},\textbf{Y})=\frac {2\cdot\sum_{ij}\min(h_{ij}, y_{ij})}{\sum_{ij}h_{ij}+\sum_{ij}y_{ij}} \enspace .
\]

\begin{figure*}[!htb]
 \centering

\includegraphics[width=16cm]{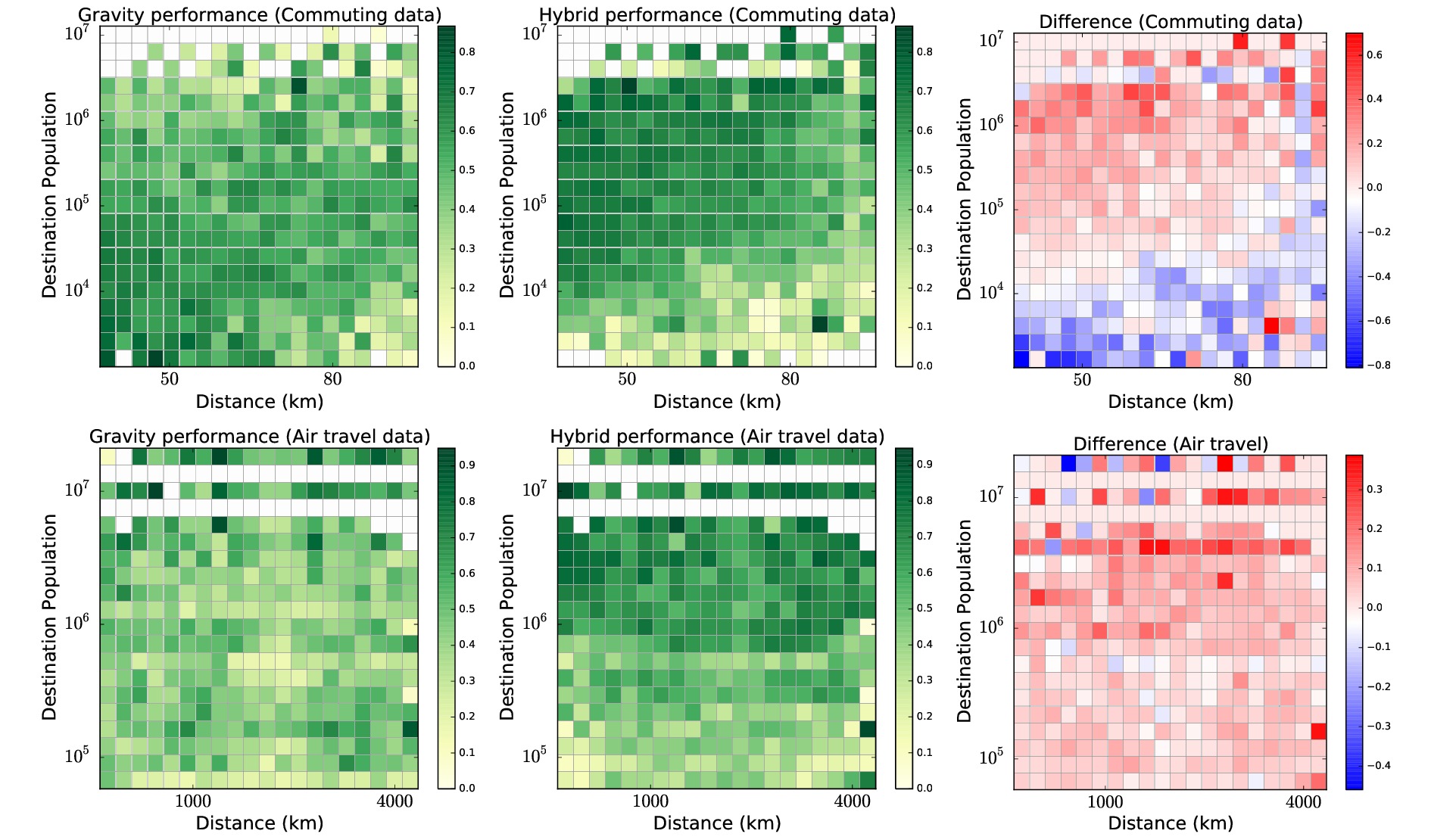}	
  \caption{{\bf Sorensen-Dice coefficient grids for the gravity model and the hybrid model.} Error of flow estimates for the gravity model (left) and the hybrid model (center), and the difference in performance between them (right) for each distance-population cell, compared to the ground-truths of commuting (top grids) and air travel (bottom grids). The Sorensen-Dice coefficients were computed following Eq. 6 in~\cite{masucci2013}.}\label{figure_sorensen}
\end{figure*}

Fig.~\ref{figure_sorensen} shows the goodness of fit in a grid of (distance, population) ranges, where each cell represents a subset of pairs of origin-destination basin filtered by distance and by destination population. In the left and central pictures, darker green colors represent a higher similarity between the predicted and real flows; the right pictures show in red those cells that were improved by the assimilation of Flickr traces, while those in blue are worse fitted when using Flickr data. We observe that in the air travel network (lower pictures) the improvement of the hybrid model is almost constant, but it is more evident for large population basins. In the commuting network, the gravity model outperforms the hybrid for small cities (up to $\approx 10,000$ inhabitants), but for larger cities the hybrid model gives better predictions. However, by improving the largest flows of people (which are usually associated to highly populated cities) the hybrid model can give a better estimation of the total amount of human flow at both resolution levels.

\bibliographystyle{unsrt} 
\bibliography{paper_nature_comm}      

\end{document}